\input harvmac
\sequentialequations

\Title{\baselineskip=14pt\vbox{\rightline{DAMTP R98/21}
\rightline{hep-th/9805014}}}
{\vbox{\centerline{Black hole absorption cross-sections and the}
       \vskip4pt\centerline{anti-de Sitter--conformal field 
       theory correspondence}}}

\centerline{Edward Teo}
\bigskip
\centerline{\sl Department of Applied Mathematics and Theoretical 
Physics, University of Cambridge,}
\centerline{\sl Silver Street, Cambridge CB3 9EW, England}
\medskip\centerline{and}
\medskip
\centerline{\sl Department of Physics, National University of 
Singapore, Singapore 119260}

\vskip 1.2in
\centerline{\bf Abstract}
\medskip
\noindent
Recent work has uncovered a correspondence between theories 
in anti-de Sitter space, and those on its boundary. This has 
important implications for black holes in string theory 
which have near-horizon AdS geometries. Using the effective 
coupling to the boundary conformal field theory, I compute the 
low-energy, $s$-wave absorption cross-sections for a minimally 
coupled scalar in the near-extremal four- and five-dimensional 
black holes. The results agree precisely with semi-classical 
gravity calculations. Agreement for fixed scalars, and for 
the BTZ black hole, is also found.

\Date{}

\nref\GT{G.W. Gibbons and P.K. Townsend, Phys. Rev. Lett. {\bf 71},
3754 (1993) {\tt hep-th/9307049}.}

\nref\GHT{G.W Gibbons, G.T. Horowitz, and P.K. Townsend, Class. 
Quantum Grav. {\bf 12}, 297 (1995) {\tt hep-th/9410073}.}

\nref\Maldacena{J. Maldacena, `The large N limit of superconformal 
field theories and supergravity,' {\tt hep-th/9711200}.}

\nref\GKP{S.S. Gubser, I.R. Klebanov, and A.M. Polyakov, `Gauge theory
correlators from non-critical string theory,' {\tt hep-th/9802109}.}

\nref\Witten{E. Witten, `Anti de Sitter space and holography,'
{\tt hep-th/9802150}.}

\nref\DMa{S.R. Das and S.D. Mathur, Nucl. Phys. {\bf B478}, 561
(1996) {\tt hep-th/9606185}.}

\nref\DMb{S.R. Das and S.D. Mathur, Nucl. Phys. {\bf B482}, 153
(1996) {\tt hep-th/9607149}.}

\nref\MSf{J. Maldacena and A. Strominger, Phys. Rev. D {\bf 55},
861 (1997) {\tt hep-th/9609026}.}

\nref\KM{I.R. Klebanov and S.D. Mathur, Nucl. Phys. {\bf B500},
115 (1997) {\tt hep-th/9701187}.}

\nref\GKT{S.S. Gubser, I.R. Klebanov, and A.A. Tseytlin,
Nucl. Phys. {\bf B499}, 217 (1997) {\tt hep-th/9703040}.}

\nref\CGKT{C.G. Callan, S.S. Gubser, I.R. Klebanov, and A.A. Tseytlin,
Nucl. Phys. {\bf B489}, 65 (1997) {\tt hep-th/9610172}.}

\nref\GubserL{S.S. Gubser, Phys. Rev. D {\bf 56}, 4984 (1997)
{\tt hep-th/9704195}.}

\nref\MSu{J. Maldacena and A. Strominger, Phys. Rev. D {\bf 56}, 
4975 (1997) {\tt hep-th/9702015}.}

\nref\Kleb{I.R. Klebanov, Nucl. Phys. {\bf B496}, 231 (1997) 
{\tt hep-th/9702076}.}

\nref\GK{S.S. Gubser and I.R. Klebanov, Phys. Lett. B {\bf 413}, 41
(1997) {\tt hep-th/9708005}.}

\nref\GHKK{S.S. Gubser, A. Hashimoto, I.R. Klebanov, and M. Krasnitz,
`Scalar absorption and the breaking of the world volume conformal
invariance,' {\tt hep-th/9803023}.}

\nref\FMMR{D.Z. Freedman, S.D. Mathur, A. Matusis, and L. Rastelli,
`Correlation functions in the CFT$_d/AdS_{d+1}$ correspondence,'
{\tt hep-th/9804058}.}

\nref\HMS{G.T. Horowitz, J.M. Maldacena, and A. Strominger, Phys.
Lett. B {\bf 383}, 151 (1996) {\tt hep-th/9603109}.}

\nref\Hyun{S. Hyun, `U-duality between three and higher dimensional 
black holes,' {\tt hep-th/9704005}.}

\nref\SS{K. Sfetsos and K. Skenderis, Nucl. Phys. {\bf B517}, 179
(1998) {\tt hep-th/9711138}.}

\nref\MSt{J. Maldacena and A. Strominger, `${\rm AdS}_3$ black holes
and a stringy exclusion principle,' {\tt hep-th/9804085}.}

\nref\BTZ{M. Ba\~nados, C. Teitelboim, and J. Zanelli, Phys. Rev. 
Lett. {\bf 69}, 1849 (1992) {\tt hep-th/9204099}.}

\nref\Satoh{Y. Satoh, `Propagation of scalars in non-extremal black 
hole and black $p$-brane geometries,' {\tt hep-th/9801125}.}

\nref\GubserA{S.S. Gubser, Phys. Rev. D {\bf 56}, 7854 (1997) 
{\tt hep-th/9706100}.}

\nref\BSS{D. Birmingham, I. Sachs, and S. Sen, Phys. Lett. B {\bf 413}, 
281 (1997) {\tt hep-th/9707188}.}
 
\nref\KK{I.R. Klebanov and M. Krasnitz, Phys. Rev. D {\bf 55},
3250 (1997) {\tt hep-th/9612051}.}

\nref\CT{M. Cveti\v c and A.A. Tseytlin, Nucl. Phys. {\bf B478},
181 (1996) {\tt hep-th/9606033}.}

\nref\GKtwo{S.S. Gubser and I.R. Klebanov, Phys. Rev. Lett. {\bf 77},
4491 (1996) {\tt hep-th/9609076}.}

\nref\BL{V. Balasubramanian and F. Larsen, `Near horizon geometry
and black holes in four dimensions,' {\tt hep-th/9802198}.}

One rather surprising result to have emerged from the study of 
certain higher-dimensional black holes and $p$-branes in string 
theory, is that their near-horizon geometries have the structure
of anti-de Sitter space \refs{\GT,\GHT}. This (and other) facts 
have lead Maldacena to conjecture that the so-called large $N$ limit 
of a conformally invariant theory in $d$ dimensions is equivalent 
to string theory on $d+1$-dimensional anti-de Sitter space 
${\rm AdS}_{d+1}$, times a sphere of constant radius \Maldacena. 
Taking this limit decouples the near-horizon region from the 
asymptotic Minkowski space, and effectively restricts one's 
attention to the former.

Subsequent work by Gubser {\it et al.\/} \GKP\ and Witten 
\Witten\ have made this relationship more precise. Their key 
insight is to put the $d$-dimensional theory on the {\it boundary\/} 
of ${\rm AdS}_{d+1}$. While the latter has a metric 
invariant under ${\rm SO}(2,d)$, its boundary has only a 
conformal structure preserved under this group. Thus, the
boundary theory has to be a conformal field theory. The link
between the bulk and boundary theories, lies in the fact that 
a dynamical field (which may be a scalar, gauge field or 
graviton) in ${\rm AdS}$, is completely specified by its boundary 
value \Witten. Supposing that the partition functions of the
two theories are equal: $Z_{\rm AdS}=Z_{\rm CFT}$, correlation 
functions of the latter can then be computed from the bulk 
theory. This is the essence of the AdS--CFT correspondence.

In \GKP, this boundary was taken to be a sphere of radius $R$ 
equal to the ${\rm AdS}$ radius of curvature, but in \Witten, 
the infinite boundary of (Euclidean) ${\rm AdS}$ space
was used. In the context of black holes however, the geometry
is only AdS near the horizon. For radii larger than $R$, 
Minkowski space is recovered. It is for this reason that I shall 
follow \GKP\ in choosing the boundary to be at $R$ rather than 
infinity. This also bypasses the problem of fields which have 
infinite asymptotic values.

In this paper, I shall show how the AdS--CFT correspondence can be 
used to derive low-energy black hole absorption cross-sections. 
Specifically, I consider minimally coupled and fixed scalars 
propagating in the near-extremal four- and five-dimensional black 
holes of string theory. These are cases which have been extensively 
studied in the past, particularly with respect to their corresponding 
D-brane configurations. In \refs{\DMa,\DMb}, it was found that 
absorption and decay rates obtained using D-branes agree exactly 
with semi-classical gravity calculations. Much work has since been 
devoted to finding an effective-string picture of D-branes that would 
model black hole scattering \refs{\MSf{--}\GubserL}.

This was carried one step further by Maldacena and Strominger \MSu, 
who argued that an effective CFT, not necessarily connected to string 
theory, is able to describe the absorption of scalars by general black 
holes. Although this effective CFT description can reproduce the 
energy dependence of the (low-energy) absorption cross-section, 
there was no known way to fix the normalization factors and other 
details, without appealing to say, string theory. The purpose of 
this paper, is to advocate that this effective theory is precisely 
the boundary CFT of the corresponding near-horizon ${\rm AdS}$ space. 
As evidence for this, I shall derive the cross-sections for the 
above-mentioned cases using the effective coupling to the boundary 
CFT, and show that they agree with semi-classical calculations. 

The D3-brane is another solution which admits a near-horizon 
${\rm AdS}$ structure \Maldacena, and the super-Yang--Mills theory 
defined on its boundary has been the subject of much recent study. 
It was shown in \GKP\ that the two-point functions obtained from 
the latter give the correct semi-classical, $s$-wave absorption 
cross-section \refs{\Kleb,\GKT,\GK}. Furthermore, the authors of 
\GHKK\ have computed higher-order corrections to it in both 
pictures. These results, together with those presented here for 
the four- and five-dimensional black holes, point to the 
central r\^ole of the AdS boundary in determining the physics 
of the scattering process.

We shall begin by reviewing the AdS--CFT correspondence, in the 
form described in \GKP. The metric for ${\rm AdS}_{d+1}$ can 
be written as
\eqn\AdS{{\rm d}s^2={R^2\over z^2}\big(-{\rm d}t^2+{\rm d}
x_1^2+\cdots{\rm d}x_{d-1}^2+{\rm d}z^2\big)\,,}
where $R$ is its radius of curvature. In this representation, 
${\rm AdS}_{d+1}$ is the upper-half space $z>0$. We take 
the boundary to be at $z=R$, enclosing the region $z>R$ of 
interest.

The action for a scalar field $\phi$, with mass $m$, in this space is
\eqn\nameless{S=\hbox{$1\over2$}\int{\rm d}^{d+1}x\sqrt{-g}
\big(g^{ab}\nabla_a\phi\nabla_b\phi+m^2\phi^2\big).}
Any solution to the field equation is completely determined 
by its behavior on the boundary, and it is possible to construct 
a Green's function relating the bulk field $\phi$ to its
boundary value $\phi_0$. $S$ can then be evaluated as a surface 
integral in terms of $\phi_0$ \GKP\ (see also the Appendix of \FMMR).

Suppose that, in the boundary CFT, $\phi_0$ is the source for a local 
field ${\cal O}$ with conformal dimension $\Delta$:
\eqn\Sint{S_{\rm int}=\int{\rm d}{\bf x}\,\phi_0({\bf x})
{\cal O}({\bf x})\,,}
where ${\bf x}\equiv(t,x_1,\dots,x_{d-1})$. The two-point function 
for ${\cal O}$ can then be deduced from $S$ to be
\eqn\correl{\langle{\cal O}({\bf x}){\cal O}({\bf y})\rangle
={2\Delta-d\over\pi^{d/2}\Delta}{\Gamma(\Delta+1)\over
\Gamma(\Delta-d/2)}{R^{2\Delta-d-1}\over|{\bf x}-{\bf y}|^{2\Delta}}\,,}
where $\Delta=d+\lambda_+$, with $\lambda_+$ being the larger 
root of $\lambda(\lambda+d)=(mR)^2$. Thus, the dimension of ${\cal O}$
is determined by the mass of $\phi$. The same result \correl\ is 
obtained from Witten's approach \Witten, provided one restores $R$ 
explicitly, and includes the extra factor found in \FMMR.

We first consider the near-extremal five-dimensional black hole of 
Type IIB string theory. The ten-dimensional form of the metric is 
\refs{\HMS,\MSf}
\eqnn\BHmetric
$$ \eqalignno{{\rm d}s^2&=(H_1H_5)^{-{1\over2}}\Bigg[-{\rm d}t^2
+{\rm d}x^2+{r_0^2\over r^2}(\cosh\sigma\,{\rm d}t+\sinh\sigma\,
{\rm d}x)^2+H_1\sum_{i=6}^9{\rm d}x_i^2\Bigg]\cr
&\quad+(H_1H_5)^{1\over2}
\Bigg[\bigg(1-{r_0^2\over r^2}\bigg)^{-1}{\rm d}r^2
+r^2{\rm d}\Omega_3^2\Bigg]\,,&\BHmetric} $$
where $x$ is periodically identified with length $2\pi R^\prime$.
$x_6,\dots,x_9$ are also compact coordinates, but play no r\^ole 
here and will be dropped. Furthermore,
\eqn\harmonic{H_1=1+{r_1^2\over r^2}\,,\qquad 
H_5=1+{r_5^2\over r^2}\,.}
$r_0$ is the extremality parameter, while $r_1$, $r_5$, and $r_k\equiv 
r_0\sinh\sigma$ are related to the charges of the black hole. The 
dilute-gas approximation that will be assumed corresponds to 
$r_0,r_k\ll r_1,r_5$ \MSf. It is convenient to define the so-called 
left- and right-temperatures:
\eqn\nameless{T_{\rm L}={r_0\over2\pi r_1r_5}{\rm e}^{\sigma},
\qquad T_{\rm R}={r_0\over2\pi r_1r_5}{\rm e}^{-\sigma},}
in which case the Hawking temperature of the black hole is given by
\eqn\temp{{2\over T_{\rm H}}={1\over T_{\rm L}}
+{1\over T_{\rm R}}\,.}
The absorption cross-section for a minimally coupled scalar in the 
$s$-wave has been calculated semi-classically to be \refs{\MSf}
\eqn\MSabs{\sigma_{\rm abs}=8\pi^3(r_1r_5)^2\omega^{-1}
T_{\rm L}T_{\rm R}\sinh\bigg({\omega\over2T_{\rm H}}\bigg)
\bigg|\Gamma\bigg(1+i{\omega\over4\pi T_{\rm L}}\bigg)
\Gamma\bigg(1+i{\omega\over4\pi T_{\rm R}}\bigg)\bigg|^2.}
This is valid if the energy $\omega$ of the incoming particle 
satisfies $\omega r_1,\omega r_5\ll1$.

To see the near-horizon AdS structure of the black hole, 
we have to take the decoupling limit as explained in \Maldacena\ 
(which is consistent with the dilute-gas approximation assumed above). 
This essentially means we ignore the 1's in the harmonic functions 
\harmonic,\foot{A different approach was used in \refs{\Hyun,\SS},
whereby the 1's are removed by a certain coordinate transformation.} 
so that \BHmetric\ becomes
\eqn\metricB{{\rm d}s^2={r^2\over R^2}(-{\rm d}t^2+{\rm d}x^2)
+{r_0^2\over R^2}(\cosh\sigma\,{\rm d}t+\sinh\sigma\,{\rm d}x)^2
+{R^2\over r^2-r_0^2}{\rm d}r^2+R^2{\rm d}\Omega_3^2\,,}
where $R^2\equiv r_1r_5$. It was explicitly shown in
\refs{\Hyun{--}\MSt} that the $(t,x,r)$ part of this metric is 
locally AdS, and in fact equivalent to the three-dimensional 
BTZ black hole \BTZ. However, we shall use a different change of 
variables, valid for large $r\gg r_0$. In this case, the second term 
in the metric \metricB\ can be neglected in comparison to the first. 
Defining the new radial coordinate $z=R^2/r$, we obtain 
\eqn\nameless{{\rm d}s^2={R^2\over z^2}(-{\rm d}t^2+{\rm d}x^2
+{\rm d}z^2)+R^2{\rm d}\Omega_3^2\,.}
This is of the form \AdS, times a sphere. The advantage of this
representation is that the boundary $z=R$ of the AdS space is 
clearly flat, although $x$ (and $t$ in the Euclidean regime) is 
periodic.

Now, it is well-known that the field equation for a minimal scalar 
$\phi$ in the $l^{\rm th}$ partial wave, reduces to the hypergeometric 
equation near the black hole horizon \refs{\KM,\MSu}. This is but 
the field equation for a scalar on ${\rm AdS}_3$, with effective 
mass-squared $m^2=l(l+2)/R^2$ \Satoh. It follows that $\phi$ 
couples to an operator ${\cal O}$ in the boundary CFT with 
conformal dimension $\Delta=2+l$. 

Let us consider the absorption of a quantum of $\phi$ with energy 
$\omega$, mediated by the interaction \Sint. We assume that 
${\cal O}(t,x)={\cal O}_+(t+x){\cal O}_-(t-x)$, where ${\cal O}_+$ 
and ${\cal O}_-$ are primary fields of dimensions $h_{\rm L}$ and 
$h_{\rm R}$ respectively. Of course, we are primarily interested 
in the case $h_{\rm L}=h_{\rm R}=\Delta/2$, but it is useful 
to be general at this stage. The absorption cross-section is given 
by \refs{\GubserA,\GK}
\eqn\nameless{\sigma_{\rm abs}={2\pi R^\prime\over2\omega}\int
{\rm d}^2x\,{\rm e}^{ip\cdot x}\big\{{\cal G}(t-i\epsilon,x)
-{\cal G}(t+i\epsilon,x)\big\}\,,}
where ${\cal G}(t,x)=\langle{\cal O}^\dagger(t,x)
{\cal O}(0,0)\rangle$ is the thermal Green's function defined in
imaginary time. The latter is determined by the singularity structure 
of ${\cal O}_\pm$ in the complex plane:
\eqn\nameless{{\cal O}_+(\bar z){\cal O}_+^\dagger(\bar w)
\sim{C_{{\cal O}_+}\over(\bar z-\bar w)^{2h_{\rm L}}}\,,\qquad
{\cal O}_-(z){\cal O}_-^\dagger(w)
\sim{C_{{\cal O}_-}\over(z-w)^{2h_{\rm R}}}\,,}
as well as the periodicity in imaginary time \refs{\MSu,\GubserA}.
It can be checked that the cross-section becomes \GubserA
\eqnn\abs
$$ \eqalignno{\sigma_{\rm abs}={\pi R^\prime C_{\cal O}\over\omega}
&{(2\pi T_{\rm L})^{2h_{\rm L}-1}(2\pi T_{\rm R})^{2h_{\rm R}-1}\over
\Gamma(2h_{\rm L})\Gamma(2h_{\rm R})}\sinh\bigg({\omega\over
2T_{\rm H}}\bigg)\times\cr&\qquad\qquad\qquad\times
\bigg|\Gamma\bigg(h_{\rm L}
+i{\omega\over4\pi T_{\rm L}}\bigg)\Gamma\bigg(h_{\rm R}
+i{\omega\over4\pi T_{\rm R}}\bigg)\bigg|^2,&\abs} $$
where $C_{\cal O}\equiv C_{{\cal O}_+}C_{{\cal O}_-}$. $T_{\rm L}$ 
and $T_{\rm R}$ are temperatures of the left- and right-moving
excitations respectively, with $T_{\rm H}$ given by \temp. The 
validity of \abs\ depends on $h_{\rm L}+h_{\rm R}$ being an integer, 
which is true for our case. The reader is referred to \GubserA\ for 
more details.

We now specialize to the massless case, corresponding
to $s$-wave scattering in the five-dimensional space-time. 
$C_{\cal O}$ can be read off from \correl\ to be $2R/\pi$, 
whence \abs\ becomes
\eqn\BTZabs{\sigma_{\rm abs}=8\pi^2RR^\prime\omega^{-1}T_{\rm L}
T_{\rm R}\sinh\bigg({\omega\over2T_{\rm H}}\bigg)
\bigg|\Gamma\bigg(1+i{\omega\over4\pi T_{\rm L}}\bigg)
\Gamma\bigg(1+i{\omega\over4\pi T_{\rm R}}\bigg)\bigg|^2.}
When $R^\prime=R$, this is precisely the cross-section for a 
minimally coupled scalar in the BTZ black hole \BSS! Recall 
that $R^\prime$ is the radius of the sixth dimension $x$ in 
\BHmetric. If we consider the BTZ black hole in its own right, 
as in \BSS, then $R^\prime$ has no particular significance and 
can be absorbed into its mass. Indeed, the angular coordinate 
$\varphi$ in the usual form of the black hole \BTZ, is related 
to $x$ by $\varphi=x/R$ \refs{\Hyun,\SS}. Demanding that it has 
canonical period $2\pi$ but corresponds to setting $R^\prime=R$.

To get the five-dimensional cross-section, we have to multiply 
\BTZabs\ by the area of the three-sphere $2\pi^2R^3$, and divide by
the circumference of the circle $2\pi R^\prime$; as obtained when 
going up to six dimensions and back down to five. It is readily 
seen that \MSabs\ is recovered. Note that $R^\prime$ cancels out.

A more exacting test would be for massive fields. Happily, such a 
case is provided by the non-minimally coupled fixed scalar, whose
$s$-wave absorption cross-section was calculated in \CGKT\ for the 
black hole \BHmetric. The effective mass-squared of this scalar 
in the AdS space is $8/R^2$, so it couples to an operator of 
conformal dimension $\Delta=4$ in the boundary CFT. The 
cross-section \abs\ becomes, after using the fact that 
$C_{\cal O}=18R^5/\pi$, and multiplying by the area 
$\pi R^3/R^\prime$, 
\eqn\nameless{\sigma_{\rm abs}=32\pi^7R^8\omega^{-1}
(T_{\rm L}T_{\rm R})^3\sinh\bigg({\omega\over2T_{\rm H}}\bigg)
\bigg|\Gamma\bigg(2+i{\omega\over4\pi T_{\rm L}}\bigg)
\Gamma\bigg(2+i{\omega\over4\pi T_{\rm R}}\bigg)\bigg|^2.}
This result was obtained in \refs{\CGKT,\KK}, by a semi-classical 
as well as D-brane calculation (with $r_1=r_5$ assumed for 
technical simplicity).

The case of the near-extremal four-dimensional black hole is 
similar. It arises from a configuration of three intersecting 
M-theory 5-branes, whose 11-dimensional metric is \CT
\eqnn\BHfour
$$ \eqalignno{{\rm d}s^2&=(H_1H_2H_3)^{-{1\over3}}\Big[-{\rm d}t^2
+{\rm d}x^2+{r_0\over r}(\cosh\sigma\,{\rm d}t+\sinh\sigma\,
{\rm d}x)^2\cr&\qquad\qquad\qquad\qquad
+H_1({\rm d}x_4^2+{\rm d}x_5^2)+H_2({\rm d}x_6^2+{\rm d}x_7^2)+
H_3({\rm d}x_8^2+{\rm d}x_9^2)\Big]\cr&\quad
+(H_1H_2H_3)^{2\over3}
\Bigg[\bigg(1-{r_0\over r}\bigg)^{-1}{\rm d}r^2
+r^2{\rm d}\Omega_2^2\Bigg]\,.&\BHfour} $$
$x$ is the $11^{\rm th}$ dimension, periodically identified with 
length $2\pi R^\prime$. $x_4,\dots,x_9$ are compact coordinates 
which we may ignore. The harmonic functions are given by
$H_i=1+r_i/r$, where $r_1$, $r_2$, $r_3$, and $r_4\equiv 
r_0\sinh^2\sigma$ characterize the charges of the black hole. As 
usual, we assume the dilute-gas approximation: $r_0,r_4\ll 
r_1,r_2,r_3$. The left- and right-temperatures of the black hole are
\eqn\nameless{T_{\rm L}={1\over4\pi}\sqrt{r_0\over r_1r_2r_3}
{\rm e}^{\sigma},\qquad T_{\rm R}={1\over4\pi}\sqrt{r_0\over 
r_1r_2r_3}{\rm e}^{-\sigma},}
and its low-energy, $s$-wave absorption cross-section is \GKtwo
\eqn\absfour{\sigma_{\rm abs}=32\pi^2r_1r_2r_3\omega^{-1}
T_{\rm L}T_{\rm R}\sinh\bigg({\omega\over2T_{\rm H}}\bigg)
\bigg|\Gamma\bigg(1+i{\omega\over4\pi T_{\rm L}}\bigg)
\Gamma\bigg(1+i{\omega\over4\pi T_{\rm R}}\bigg)\bigg|^2,}
for the case of minimal scalars.

When the decoupling limit is taken, \BHfour\ reduces to the BTZ black 
hole with radius of curvature $R=2(r_1r_2r_3)^{1\over3}$, times
a two-sphere of radius $R/2$ \refs{\Hyun,\SS,\BL}. The effective 
CFT calculation is thus identical to the five-dimensional case, 
except that we have to multiply by $R^2/2R^\prime$ in going from three
to four dimensions \BL. Indeed, \absfour\ is recovered from \BTZabs.

The fixed scalar in this black hole again has an effective 
mass-squared $8/R^2$ \CGKT, and so $\Delta=4$. Repeating 
the by-now familiar procedure, we find the CFT predicts a 
cross-section
\eqn\nameless{\sigma_{\rm abs}=16\pi^6R^7\omega^{-1}
(T_{\rm L}T_{\rm R})^3\sinh\bigg({\omega\over2T_{\rm H}}\bigg)
\bigg|\Gamma\bigg(2+i{\omega\over4\pi T_{\rm L}}\bigg)
\Gamma\bigg(2+i{\omega\over4\pi T_{\rm R}}\bigg)\bigg|^2,}
in agreement with the result of \KK\ (where $r_1=r_2=r_3$ was
assumed).

We have therefore seen that the boundary CFT is able to reproduce 
the low-energy absorption cross-sections of black holes in a
few important cases. The physical interpretation of this should 
be clear \GKP: particles coming in from infinity would encounter 
a geometry that approximates AdS space from radius $R$ onwards, 
and any subsequent evolution is encoded in this boundary.
What is perhaps less obvious is the r\^ole of the sixth (or 
$11^{\rm th}$) dimension, which provides the effective length-scale 
of the CFT. Although $R^\prime$ disappears in the final 
cross-sections, it may still be relevant in the ten- (or 11-)
dimensional context. Ref.~\HMS\ contains a discussion on how 
$R^\prime$ affects the physics from the D-brane point of view.

It would be of interest to extend these results to higher partial
waves. Since they are not spherically symmetric, there would be 
a non-trivial variation over the two- or three-sphere which must 
be taken into account.\foot{Nevertheless, a quick calculation shows
that the boundary CFT above gives an answer which is tantalizingly
close to the semi-classical result \refs{\KM,\MSu,\GubserL}.}
Also notice that we have not used any knowledge from string theory in 
the above calculations; all that was assumed was a scalar propagating
in the space-time. A question which then follows is, how universal
are these results? Indeed, it was pointed out in \Satoh\ that the 
minimal scalar absorption cross-sections for black holes and 
$p$-branes in diverse dimensions share the CFT structure of \abs. 
The near-horizon AdS geometry of these objects should probably 
explain it.

\listrefs
\bye